Civic Engagement among Community Computer Network Users:

Trend or Phase?

Andrea L. Kavanaugh
Virginia Polytechnic Institute & State University
Blacksburg, VA 24061-0506
Kavan@vt.edu

TPRC 2001

**DRAFT**

October 26, 2001

**Abstract**

People use communication and information tools to meet existing needs and purposes. In Fischer's (1992) words, 'Americans used the telephone to enhance the ways of life to which they were already committed.' As such, media have 'dual effects' (Pool, 1982) depending on an individual's prior behavior, needs and interests. This paper brings evidence to bear on the question of the long-term effects of Internet diffusion on civic engagement in geographic communities. It draws on findings from survey data collected in four U.S. towns and cities in fall 2000 where community computer networking is established. The study shows that the evidence in these four sites is consistent with earlier findings in Blacksburg, Virginia (Kavanaugh, 2000; Patterson and Kavanaugh, 2001; Kavanaugh and Patterson, 2001) and other studies of early innovation adopters (Rogers, 1983; Kohut, 1999; Valente, 1995, among others). The data indicate that early adopters of the Internet are more likely to engage in civic activities and to have higher levels of community involvement than later adopters. Further, early adopters are more likely to use the Internet to increase their community involvement, political participation and connectivity with like-minded people. Later adopters in all four sites show less involvement in their local community and less interest in political activity and information, particularly online. The findings summarized in this paper reinforce those of the Kohut (1999) study showing that later adopters are less civic minded and more interested than early adopters in consumer and commercial applications, such as shopping and entertainment. The results reported in this paper lend weight to the argument that increases in civic engagement and community involvement are due primarily to the behavior of early adopters, making such increases a phase, not a trend. As later adopters come on line, use of the Internet for community involvement or civic engagement decreases. In the long term, Internet access will have only a modest effect on community involvement and civic engagement in geographic communities.

**Models of the Local Community**

In trying to explain the impact of computer networking on social ties and participation involvement in geographic communities, early models of local community provide a useful point of departure. A "linear development model" explains social ties and participation in community primarily by differences in population size and density. This model has roots in Toennies (1887) and Wirth (1938). This model predicts a decline in community, as urban ways of life continuously weaken social networks and bonds.

Another model of local community arising from the Chicago school of urban sociology, a "systemic model" of local community counters that In this model, length of residence, social status and life-cycle stage are key predictors of local social ties and community participation (Axelrod, 1956; Wilensky, 1961; Laumann, 1973; Kasarda and Janowitz, 1974). Higher status individuals tend to have smaller proportions of their friends and relatives residing within their own communities and fewer relatives living nearby due to their greater mobility. For the same reason, they tend to belong to more formal organizations in the community. Due to their greater mobility (or rootlessness and weaker local social networks), higher status individuals tend to rely more heavily on formal or "secondary" social networks through local formal organizations. At the same time, higher status persons have a greater stake in the community and a higher ability to articulate their interests. Therefore, involvement in the affairs of the local community is influenced most by a person's position in the social structure (Kasarda and Janowitz, 1974). Some more recent studies of community attachment (Stamm, 1985) and involvement (Rothenbuhler, 1991), concur that social status (measured by education, income and occupation) is an important factor. Social status (class) and family cycle are two of the most powerful variables working on structure and content of a person's social network (Bernard, 1973, p. 53). The higher levels of formal education among persons of higher social status often reflect a natural tendency or predisposition toward leadership - a characteristic of early adopters of an innovation.

**Early Adopters and Civic Engagement**

Diffusion theory claims that the characteristics of early adopters of an innovation are distinct from later and non-adopters of an innovation. Moreover, these characteristics are consistently associated with higher levels of civic engagement and community involvement. Early adopters of innovations are distinguished from later adopters and non-adopters by higher levels of education and social status, media exposure, cosmopolitanness and empathy (Rogers, 1983). Diffusion studies of computer use show that formal education is the single variable most consistently associated with the adoption of personal computing; social status is also a consistent predictor (Dutton, Rogers and Jun, 1987). Early adopters are opinion leaders, what Markus (1990) calls 'high resource individuals', who are sought after for their knowledge and expertise (Katz and Lazarsfeld, 1964).

Diffusion researchers have established the importance of opinion leadership and social networks in understanding how an innovation diffuses through a social system (Coleman et al, 1957; Kling and Gerson, 1977; Rogers 1983; Markus, 1990; Valente,

1995, Patterson and Kavanaugh 2001). Opinion leaders share their experiences about an innovation with members of their social networks, thereby making the social networks of early adopters critical to the diffusion process.

**Towards a Model of the Networked Community:**

This paper draws on the systemic model of community and diffusion of innovation studies, to explain the effects of computer networking on local social ties and community participation. This paper presents an examination of the variables found to be significant in the systemic model of community, together with variables significant in Internet adoption and diffusion. It evaluates several hypotheses to determine whether early adopters of community computer networking show classic characteristics of early adopters, including high levels of education and social status, and high levels of civic engagement and community involvement. Further, the study tests whether early adopters are different from later adopters in terms of their use of the Internet for strengthening local social ties and community participation. The hypotheses tested in this study are:

> *H1: Persons of higher social status and longer residence in the community will have higher levels of interest and involvement in local affairs*
>
> *H2: Persons of higher social status will adopt and use the community computer network earlier than others in the community.*

> *H3: Early adopters will use the computer network to facilitate their interest and involvement in local affairs.*

We expect early adopters of community computer networking to have higher levels of education and social status based on diffusion theory and research, as well as prior studies on community. As opinion leaders, we expect early adopters to have higher levels of community involvement and civic engagement, and to use the Internet to increase their involvement. The confirmation of these hypotheses would suggest that a decline in the use of community computer networks for pursuing interest and involvement in local affairs is a trend not phase in the diffusion of community computer networks.

**Research Methods**

The data reported in this paper were collected using a random sample mail survey designed by the author in collaboration with co-principal investigators on the project.[i] The sample for this survey is a random household sample produced by a private company, Survey Sample, Incorporated (SSI). SSI created the sample in September 2000 from telephone directory records, post office records, and DMV data, updated on an ongoing basis. The sampling frame is comprised of all households within the zip code areas associated with four sites in the U.S. where community computer networks are established.   These sites are: the city limits of the city of Pittsburgh and Allegheny County, the town of Blacksburg and Montgomery County, the twin cities of Champaign and Urbana, and the city of Seattle.  In order to have a final sample of about 1,000 per site, SSI drew up an initial sample of 1,250 households, with the expectation that they were 80% deliverable. SSI used filtered out any postal address or zip code associated with a campus dormitory, as well as any addresses designated "rural delivery" with no street addresses.

Measures social status are education and income. Measures for community involvement are drawn from the set of Rothenbuhler (1991) questions. These are: how often the respondent keeps up with local news, gets together with others who know what is going on in the community, has ideas for improving things in the community, and works to bring about change in the community.

The random sample population in each of the four sites received an advance letter co-signed by a local government representative and the director of the local community network inviting them to participate in the study. About one week later, they received the mail survey, followed by two postcard reminders.  One of the limitations of the study is that the response rate is low at each site (between 150 to 200 out of about 1,000 surveys deliverable per site), bringing the total N to under 600 (15%).  Hypotheses were tested with regression models, T-tests, Tukey's Studentized Range tests, and correlations using the Statistical Package for the Social Sciences (10.0 version).

**Results**

**Hypothesis 1**, which predicts that persons with higher social status and longer residence in the community will have higher levels of interest and involvement in local affairs, is confirmed. Persons with higher social status show higher levels of offline community involvement than later adopters and non-adopters (Table 1a, Social Status, Length of Residence and Community Involvement). They also are more likely persons of lower social status to engage in civic and political processes and activities (Table 1b, Social Status and Community Involvement). People with higher social status also have higher levels of newspaper readership, and home ownership, measures that are often used to indicate community involvement and attachment.

Hypothesis 2, which predicts that person of higher social status will be early adopters of the community computer network, is confirmed. The more education people have ($p<.000$), and the higher their income ($p<.003$), the more likely they are to have used the Internet for a longer period of time (Table 2).

Hypothesis 3, which predicts that early adopters with longer residence will use the community computer network to facilitate their interest and involvement in local affairs, is confirmed. Early adopters will use the Internet to act on their interest in local issues, and to increase their community involvement (Table 3). Early adopters are more likely than later adopters to use the Internet for the same types of political activities online that they conduct off line.

**Discussion**

There are many reasons why Americans do not participate in political processes, discussions and institutions. Convenient access to communication tools and information sources appears to be a minor reason. This reason probably applies primarily to people who are already engaged in civic affairs or are 'predisposed' to become more involved. They are interested in civic issues and community life, but suffer from increasing demands on their time and attention. The convenience of computer networking for obtaining and exchanging information, and for communicating with others, may be a crucial factor in increasing participation among these individuals.

The data presented in this paper show that early adopters of the Internet have higher levels of education and social status than later adopters and non-adopters. Moreover, early adopters are generally more likely to engage in civic activities than later adopters, particularly online civic activities. Finally, early adopters are more likely to use the Internet to increase their involvement in the local community and to engage in civic life than later adopters. Increases in civic engagement and community involvement among Internet adopters are, therefore, likely to be a phase, not a trend. If this argument holds, we can expect that as later adopters come on line, use of the Internet for community involvement or civic engagement will decrease among the total population of Internet users. In the long term, Internet access will have only a modest effect on community involvement and civic engagement in geographic communities, and will be limited to those individuals (early adopters) with higher levels of education and social status, who have traditionally been more actively involved in their local communities.

**Tables**

**Table 1a High Social Status and Long Residence by Community Involvement**

|  | N | Sum of Squares | df | F | Sig |
|---|---|---|---|---|---|
| **Community Involvement** | 280 | .19.030 | 5 | 10.689 | .000 |

**Table 1b Social Status by Community Involvement**

|  | N | Sig. (2-tailed) | Community Involvement (Pearson Correlation) |
|---|---|---|---|
| **Education** | 407 | .002 | .153** |
| **Income** | 377 | .000 | .203** |

** Correlation is significant at the 0.01 level

**Table 2: Income and Education by Length of Internet Use**

|  | Std. Dev. | N | Sig.(2-tailed) | Years of Use |
|---|---|---|---|---|
| **Education** | .8895 | 390 | .000 | .269** |
| **Income** | 3.27 | 360 | .003 | .156** |

** Correlation is significant at the 0.01 level

Table 3
Descriptive Statistics and ANOVA Tests for
Differences by Length of Internet Use
Early vs Late Adopters and Internet Use
N=191 (Early Adopters); N= 125 (Later Adopters)

|  | Early Adopter 4+ years | | Later Adopter 0-3 years | | |
|---|---|---|---|---|---|
| **Variable** | **Mean** | **SD** | **Mean** | **SD** | **Significance Test** |
| Community Involvement | | | | | |
|    Work to bring change | 2.31 | .85 | 2.06 | 1.03 | F=6.793, p<.01 |
|    Distribute Political Info | 1.66 | .99 | 1.44 | .82 | F=4.161, p<.05 |
|    Attend Civic Org Mtgs | 2.05 | .378 | 1.93 | .356 | F=6.961, p<.01 |
| | | | | | |
| Online Political Activity | | | | | |
|    Online petition | 1.49 | .95 | 1.27 | .79 | F=4.446, p<.05 |
|    Email Govt Official | 1.96 | 1.12 | 1.58 | .93 | F=10.309, p<.001 |
|    Discuss Issues Online | 1.71 | 1.13 | 1.35 | .77 | F=10.309, p<.05 |
|    Volunteer activity | 1.28 | .70 | 1.12 | .49 | F=4.669, p<.05 |
|    Obtain Political Info | 2.62 | 1.41 | 1.88 | 1.23 | F=22.605, p<.001 |
|    Visit Political Website | 2.22 | 1.29 | 1.66 | 1.08 | F=16.279, p<.001 |
|    Visit Local Civic Site | 3.09 | 1.18 | 2.75 | 1.13 | F=5.822, p<.05 |
| | | | | | |
| Increased Involvement since Net Access | | | | | |
|    Involved w/Issues | 2.42 | .545 | 2.27 | .483 | F=5.533, p<.05 |
|    Involved w/Community | 2.07 | .407 | 1.92 | .303 | F=13.228, p<.001 |
|    Connected w/People | 2.45 | .578 | 2.23 | .527 | F=11.515, p<.001 |

___

---

[i] Co-principal investigators on this project are Joseph Schmitz, and Scott Patterson.